\documentclass[prd,superscriptaddress,twocolumn,showpacs,dvips]{revtex4}
\usepackage{feynmp}
\usepackage{amssymb}
\usepackage{epsfig}
\usepackage{graphicx}
\usepackage{subfigure}
\usepackage{hyperref}
\usepackage{bbm}
\usepackage{color}

\begin{document}

\title{Renormalization of fermion velocity in finite temperature QED$_{3}$}

\author{Jing-Rong Wang}
\affiliation{High Magnetic Field Laboratory, Hefei Institutes of
Physical Science, Chinese Academy of Sciences, Hefei, Anhui 230031,
P. R. China}

\author{Guo-Zhu Liu}
\affiliation{Department of Modern Physics, University of Science and
Technology of China, Hefei, Anhui 230026, P. R. China}

\author{Chang-Jin Zhang}
\affiliation{High Magnetic Field Laboratory, Hefei Institutes of
Physical Science, Chinese Academy of Sciences, Hefei, Anhui 230031,
P. R. China} \affiliation{Collaborative Innovation Center of
Advanced Microstructures, Nanjing University, Nanjing 210093, P. R.
China}

\begin{abstract}
At zero temperature, the Lorentz invariance is strictly preserved in
three-dimensional quantum electrodynamics. This property ensures
that the velocity of massless fermions is not renormalized by the
gauge interaction. At finite temperature, however, the Lorentz
invariance is explicitly broken by the thermal fluctuation. The
longitudinal component of gauge interaction becomes short-ranged due
to thermal screening, whereas the transverse component remains
long-ranged because of local gauge invariance. The transverse gauge
interaction leads to singular corrections to the fermion self-energy
and thus results in an unusual renormalization of the fermion
velocity. We calculate the renormalized fermion velocity $v^R(p_0,
\mathbf{p}, T)$ by employing a renormalization group analysis, and
discuss the influence of the anomalous dimension $\eta_n$ on the
fermion specific heat.
\end{abstract}


\maketitle


Four-dimensional quantum electrodynamics (QED$_4$) can describe the
electromagnetic interaction with very high precision after
eliminating ultraviolet divergences by means of renormalization
method. Different from QED$_{4}$, (2+1)-dimensional QED of massless
fermions, dubbed QED$_{3}$, is superrenormalizable and does not
contain any ultraviolet divergence. However, extensive
investigations have showed that QED$_{3}$ exhibits a series of
nontrivial low-energy properties, such as dynamical chiral symmetry
breaking (DCSB) \cite{Pisarski84, Appelquist86, Appelquist88,
Nash89, Atkinson90, Curtis90, Pennington91, Curtis92, Maris96,
Fisher04, Bashir08, Bashir09, Goecke09, Lo11, Braun14, Roberts94,
Appelquist04}, asymptotic freedom \cite{Appelquist86}, and weak
confinement \cite{Burden92, Maris95, Bashir08}. It thus turns out
that QED$_{3}$ is more similar to four-dimensional quantum
chromodynamics (QCD$_{4}$) than QED$_4$. For this reason, QED$_{3}$
is widely regarded as a toy model of QCD$_{4}$ in high-energy
physics. On the other hand, in the past decades QED$_{3}$ has proven
to be an effective low-energy field theory for several important
condensed-matter systems, including high-$T_c$ cuprate
superconductors \cite{Lee06, Affleck, Kim97, Kim99, Rantner, Franz,
Herbut, Liu02, Liu03}, spin-$1/2$ Kagome spin liquid \cite{Ran07,
Hermele08}, graphene \cite{Gusynin04, Gusynin07, Raya08}, certain
quantum critical systems \cite{Klebanov, Lu14}, surface states of
some bulk topological insulators \cite{Metlitski15, Wang15,
Mross15}.

Appelquist \emph{et al.} analyzed the Dyson-Schwinger equation (DSE)
of fermion mass in zero-$T$ QED$_3$ and revealed that the massless
fermions can acquire a finite dynamical mass, which induces DCSB,
when the fermion flavor is below some threshold, i.e., $N < N_c$
\cite{Appelquist88}. Most existing analytical and numerical
calculations \cite{Nash89, Atkinson90, Curtis90, Pennington91,
Curtis92, Maris96, Fisher04, Roberts94, Appelquist04} agree that
$N_c \approx 3.5$ at zero $T$. This problem is not only interesting
in its own right, but of practical importance since QED$_{3}$ has
wide applications in condensed-matter physics \cite{Lee06, Affleck,
Kim97, Kim99, Rantner, Franz, Herbut, Liu02,Liu03, Ran07, Hermele08,
Gusynin04, Gusynin07, Raya08, Klebanov, Lu14, Metlitski15,Wang15,
Mross15}. In particular, it has been demonstrated \cite{Affleck,
Kim97, Kim99, Liu02, Franz, Herbut} that DCSB leads to the formation
of quantum antiferromagnetism. Dynamical mass generation at finite
temperature in QED$_{3}$ is also an interesting, and meanwhile very
complicated, issue that has been investigated for over two decades
\cite{Dorey92, Lee98,Triantaphyllou,Feng13, Feng14, Lo14, WangLiuZhang15}.

If the fermion flavor is large, say $N \geq 4$, no DCSB takes place
and the Dirac fermions are still massless despite of the presence of
strong gauge interaction. However, QED$_3$ is still highly
nontrivial in its massless phase, because the gauge interaction can
lead to unusual, non-Fermi liquid like behaviors of fermions
\cite{Kim97, Kim99, WangLiu10A, WangLiu10B}. These non-Fermi liquid
behaviors may be of important relevance to the low-energy physics of
high-$T_c$ cuprate superconductors \cite{Kim97, Kim99, Lee06} and
other strongly correlated systems.

When studying the non-Fermi liquid behaviors of massless fermions,
an important role is known to be played by the fermion velocity $v$,
which enters into many observable quantities of massless fermions,
such as specific heat \cite{Kim97,Vafek07, Xu08, JWangLiu12,
Liu2015} and thermal conductivity \cite{Fritz09}. An interesting
property is that the constant velocity can be renormalized by
various interactions and then exhibits unusual momentum dependence.
For instance, it is known that the low-energy elementary excitations
of graphene, a single layer of carbon atoms, are massless Dirac
fermions \cite{Kotov}. The fermion velocity in graphene is certainly
a constant in the non-interacting limit, and its bare value is
roughly $c/300$ with $c$ being the speed of light in vacuum
\cite{Kotov}. However, extensive renormalization group (RG) analysis
\cite{Gonzalez94, Gonzalez99, Son07, Kotov} have showed that the
fermion velocity can be enhanced by the unscreened long-range
Coulomb interaction. In the lowest energy limit, the velocity flows
to very large values and the fermion dispersion is thus
substantially modified. Remarkably, the predicted nearly divergence
of the renormalized velocity has already been confirmed in recent
experiments \cite{Elias11, Yu13, Siegel12}. Another notable example
is the effective QED$_3$ theory of high-$T_c$ superconductors
\cite{Lee06, Kim97, Kim99}, which contains only the transverse part
of the U(1) gauge interaction. Moreover, near the nematic quantum
critical point in high-$T_c$ superconductors, massless Dirac
fermions interact strongly with the quantum fluctuation of nematic
order parameter \cite{Huh08, Wang11, Liu12, She15}. In both cases,
the fermion velocity receives singular corrections and is driven to
vanish in the low-energy region \cite{Lee06, Kim97, Kim99, Huh08,
Wang11, Liu12, She15}, which in turn results in non-Fermi liquid
behaviors \cite{Lee06, Kim97, Kim99, Liu2015}.

In this paper, we study the renormalization of fermion velocity due
to U(1) gauge interaction in QED$_3$. Whether the velocity is
renormalized depends crucially on the temperature of the system. At
zero temperature, the Lorentz invariance of QED$_3$ is certainly
reserved, and thus there is no interaction corrections to the
fermion velocity. In this case, fermion velocity is always a
constant. At finite temperature, however, the Lorentz invariance is
explicitly broken by thermal fluctuations \cite{Dorey92}. As a
result, the longitudinal and transverse parts of the gauge
interaction are no longer identical, and the fermion velocity may
flow with varying energy and momenta. It is interesting to ask two
questions: How the fermion velocity is renormalized by the gauge
interaction at finite temperature? How the physical quantities of
Dirac fermions, such as the specific heat, are influenced by the
renormalzied velocity? In this article, we will study these two
questions.

We study this problem and calculate the renormalized fermion
velocity $v^R(p_0, \mathbf{p}, T)$ by means of renormalization group
method. We show that the velocity exhibits a power law
dependence on momentum $|\mathbf{p}|$ under the energy scale $T$,
\begin{eqnarray}
v^R(p_0, \mathbf{p}, T) =
\left(\frac{|\mathbf{p}|}{T}\right)^{\eta},
\end{eqnarray}
where the anomalous dimension $\eta$ are functions of both $T$ and
energy $p_0$, namely $\eta \equiv \eta(p_0,T)$. We then study the
impact of the renormalized fermion velocity on the specific heat
of massless Dirac fermions.

The Lagrangian density for QED$_{3}$ with $N$ flavors of massless
Dirac fermions is given by
\begin{eqnarray}
\mathcal{L}=\sum_{i=1}^{N}\bar{\psi}_{i}\left(i\partial\!\!\!\slash
+ eA\!\!\!\slash \right)\psi_{i}
-\frac{1}{4}F_{\mu\nu}^{2},\label{Eq:Langrangian}
\end{eqnarray}
where $F_{\mu\nu} = \partial_{\mu}A_{\nu} -
\partial_{\nu}A_{\mu}$. The fermion is described by a four-component
spinor $\psi$, whose conjugate is $\bar{\psi} =
\psi^{\dag}\gamma_{0}$. The gamma matrices are defined as
$(\gamma_{0},\gamma_{1},\gamma_{2}) =
(i\sigma_{3},i\sigma_{1},i\sigma_{2})\otimes\sigma_{3}$, which
satisfy the Clifford algebra $\{\gamma_{\mu},\gamma_{\nu}\} =
2g_{\mu\nu}$ with $g_{\mu\nu} = \mathrm{diag}(-1,-1,-1)$. In (2+1)
dimensions, there are two chiral matrices, denoted by
$\gamma_{3}=I_{2\times2}\otimes\sigma_{1}$ and
$\gamma_{5}=I_{2\times2}\otimes\sigma_{2}$ respectively, that
anti-commute with $\gamma_{0,1,2}$. This Lagrangian respects a
continuous $U(2N)$ chiral symmetry $\psi \rightarrow e^{i\theta
\gamma_{3,5}}\psi$, where $\theta$ is an arbitrary constant. Once
fermions become massive, the $U(2N)$ chiral symmetry is broken down
to $U(N)\times U(N)$. Here, we consider a general very large $N$,
which implies the absence of DCSB, and perform perturbative
expansion in powers of $1/N$. For simplicity, we work in units with
$\hbar = k_{B} = 1$.

The fermions have a constant velocity $v$, which appears in the
covariant derivative in the form $\partial\!\!\!\slash = \gamma_0
\partial_0 + v \mathbf{\gamma}\cdot \mathbf{\nabla}$. In most previous
studies, it is assumed that $v \equiv 1$. At zero $T$, this
assumption is perfectly good and not affected by the gauge
interaction since the Lorentz invariance is absolutely satisfied. To
see this, we write the fermion propagator as
\begin{eqnarray}
G_0(k_0,\mathbf{k}) = \frac{1}{\gamma_0 k_0 + v\mathbf{\gamma}\cdot
\mathbf{k}}.
\end{eqnarray}
The self-energy corrections due to gauge interaction is generically
expressed as
\begin{eqnarray}
\Sigma(k_0,\mathbf{k}) = A_0(k_0, \mathbf{k})\gamma_0 k_0 +
A_\mathrm{s}(k_0, \mathbf{k})v \mathbf{\gamma}\cdot \mathbf{k}.
\end{eqnarray}
Here,
$A_{0,\mathrm{s}}(k_0, \mathbf{k})$ are the temporal and spatial
components of the wave function renormalization respectively. At $T
= 0$, the Lorentz invariance ensures that
\begin{eqnarray}
A_0(k_0, \mathbf{k}) = A_\mathrm{s}(k_0, \mathbf{k}) \equiv A(k),
\end{eqnarray}
thus the dressed fermion propagator has the form
\begin{eqnarray}
G(k_0,\mathbf{k}) = \frac{1}{\left[1+A(k)\right]\left(\gamma_0 k_0 +
v \mathbf{\gamma}\cdot \mathbf{k}\right)}.
\end{eqnarray}
Clearly, the velocity $v$ remains a constant and does not receive
any interaction corrections. It is therefore safe to set $v \equiv
1$. However, this is no longer true when the Lorentz invariance is
explicitly broken at finite $T$. Once the Lorentz invariance is
broken, we have $A_0(k_0, \mathbf{k}) \neq A_\mathrm{s}(k_0,
\mathbf{k})$. The difference $\Delta A = A_\mathrm{s}(k_0,
\mathbf{k}) - A_0(k_0, \mathbf{k})$ represents the interaction
correction to the fermion velocity $v$, which then becomes a
function of energy-momenta and temperature, i.e., $v \rightarrow
v^R(p_0, \mathbf{p}, T)$. We now calculate the function $v^R(p_0,
\mathbf{p}, T)$ by means of RG method.

We will work in the standard Mastubara formalism for finite $T$
quantum field theory, and assume the fermion energy to be of the
form $k_{0}=(2n+1)\pi T$ with $n$ being an integer. Including the
correction of the polarizations, the effective propagator of gauge
boson now becomes
\begin{eqnarray}
\Delta_{\mu\nu}(q_{0},\mathbf{q}) &=&
\frac{A_{\mu\nu}}{\frac{q_{0}^2}{v^{2}} +
\mathbf{q}^{2}+\Pi_{A}(q_{0},\mathbf{q})} \nonumber \\
&+& \frac{B_{\mu\nu}}{\frac{q_{0}^{2}}{v^{2}} +
\mathbf{q}^2+\Pi_{B}(q_{0},\mathbf{q})},
\end{eqnarray}
where $q_0 = 2m\pi T$ with $m$ being an integer. The two tensors
$A_{\mu\nu}$ and $B_{\mu\nu}$ are defined as
\begin{eqnarray}
A_{\mu\nu} &=& \left(\delta_{\mu0} - \frac{q_{\mu}q_{0}}{q^2}\right)
\frac{q^2}{v^{2}\mathbf{q}^2}\left(\delta_{0\nu} -
\frac{q_{0}q_{\nu}}{q^2}\right), \\
B_{\mu\nu} &=& \delta_{\mu i}\left(\delta_{ij} -
\frac{q_{i}q_{j}}{\mathbf{q}^{2}}\right)\delta_{j\nu}.
\end{eqnarray}
It is easy to verify that $A_{\mu\nu}$ and $B_{\mu\nu}$ are
orthogonal and satisfy
\begin{eqnarray}
A_{\mu\nu}+B_{\mu\nu}=\delta_{\mu\nu}-\frac{q_{\mu}q_{\nu}}{q^2}.
\end{eqnarray}
The polarizations  $\Pi_{A}$ and $\Pi_{B}$ are defined by
\begin{eqnarray}
\Pi_{A} = \frac{q^{2}}{v^{2}\mathbf{q}^{2}}\Pi_{00},\qquad \Pi_{B} =
\Pi_{ii}-\frac{q_{0}^{2}}{v^{2}\mathbf{q}^{2}}\Pi_{00}£¬
\end{eqnarray}
where
\begin{figure}[htbp]
\center
\includegraphics[width=3.2in]{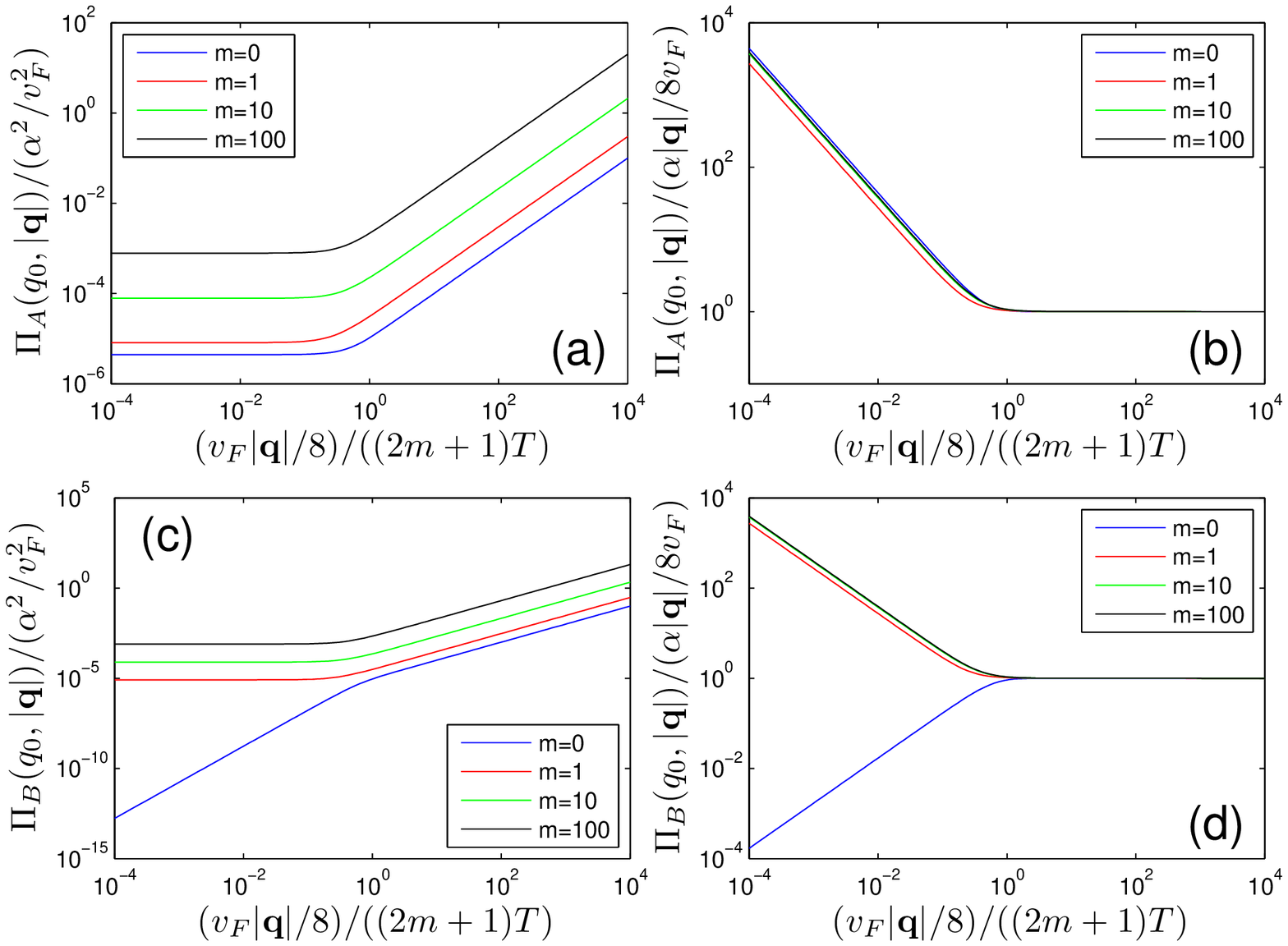}
\caption{Polarizations $\Pi_{A}(q_{0},|\mathbf{q}|)$ and
$\Pi_{B}(q_{0},|\mathbf{q}|)$ for different frequencies. Here, $q_0
= 2m\pi T$ and $T$ is chosen as
$T/\alpha=10^{-5}$.}\label{Fig:PolaFiniteFreA}
\end{figure}
\begin{eqnarray}
\Pi_{00} &=& \frac{\alpha}{\beta}
\sum_{k_0}\int\frac{d^{2}\mathbf{k}}{(2\pi)^{2}} \mathrm{Tr}
\left[G(k_{0},\mathbf{k})\gamma_{0}G(k_{0} +
q_{0},\mathbf{k}+\mathbf{q})\gamma_{0}\right], \nonumber \\
\Pi_{ii} &=& \frac{\alpha}{\beta}\sum_{k_0} \int
\frac{d^{2}\mathbf{k}}{(2\pi)^{2}}\mathrm{Tr}
\left[G(k_{0},\mathbf{k})\gamma_{i}G(k_{0} +
q_{0},\mathbf{k}+\mathbf{q})\gamma_{i}\right],\nonumber
\end{eqnarray}
with $\alpha=Ne^2$. As usual \cite{Appelquist88}, the parameter
$\alpha$ is kept fixed as $N \rightarrow +\infty$. Employing the
method utilized in Ref.~\cite{Dorey92}, one can obtain the following
expressions:
\begin{eqnarray}
\Pi_{A}=\Pi_{3},\qquad \Pi_{B}=\Pi_{1}+\Pi_{2}
\end{eqnarray}
where
\begin{eqnarray}
\Pi_{1}&=&\frac{\alpha}{2\pi v^{2}}\int_{0}^{1}dx
\frac{\chi\sinh\left(\frac{\chi}{T}\right)}{\cosh^2
\left(\frac{\chi}{2T}\right) - \sin^2\left(\frac{x\
q_{0}}{2T}\right)},\nonumber \\
\Pi_{2} &=& \frac{\alpha q_{0}}{4\pi v^{2}} \int_{0}^{1}dx
\frac{(1-2x)\sin\left(\frac{xq_{0}}{T}\right)}{\cosh^2
\left(\frac{\chi}{2T}\right) - \sin^2 \left(\frac{x
q_0}{2T}\right)}, \nonumber \\
\Pi_{3} &=& \frac{\alpha T}{\pi v^{2}}\int_{0}^{1}dx \nonumber \\
&& \times \ln\left(4\left[\cosh^2\left(\frac{\chi}{2T}\right) -
\sin^2\left(\frac{x q_{0}}{2T}\right)\right]\right),\nonumber
\end{eqnarray}
\begin{figure}[htbp]
\includegraphics[width=3.2in]{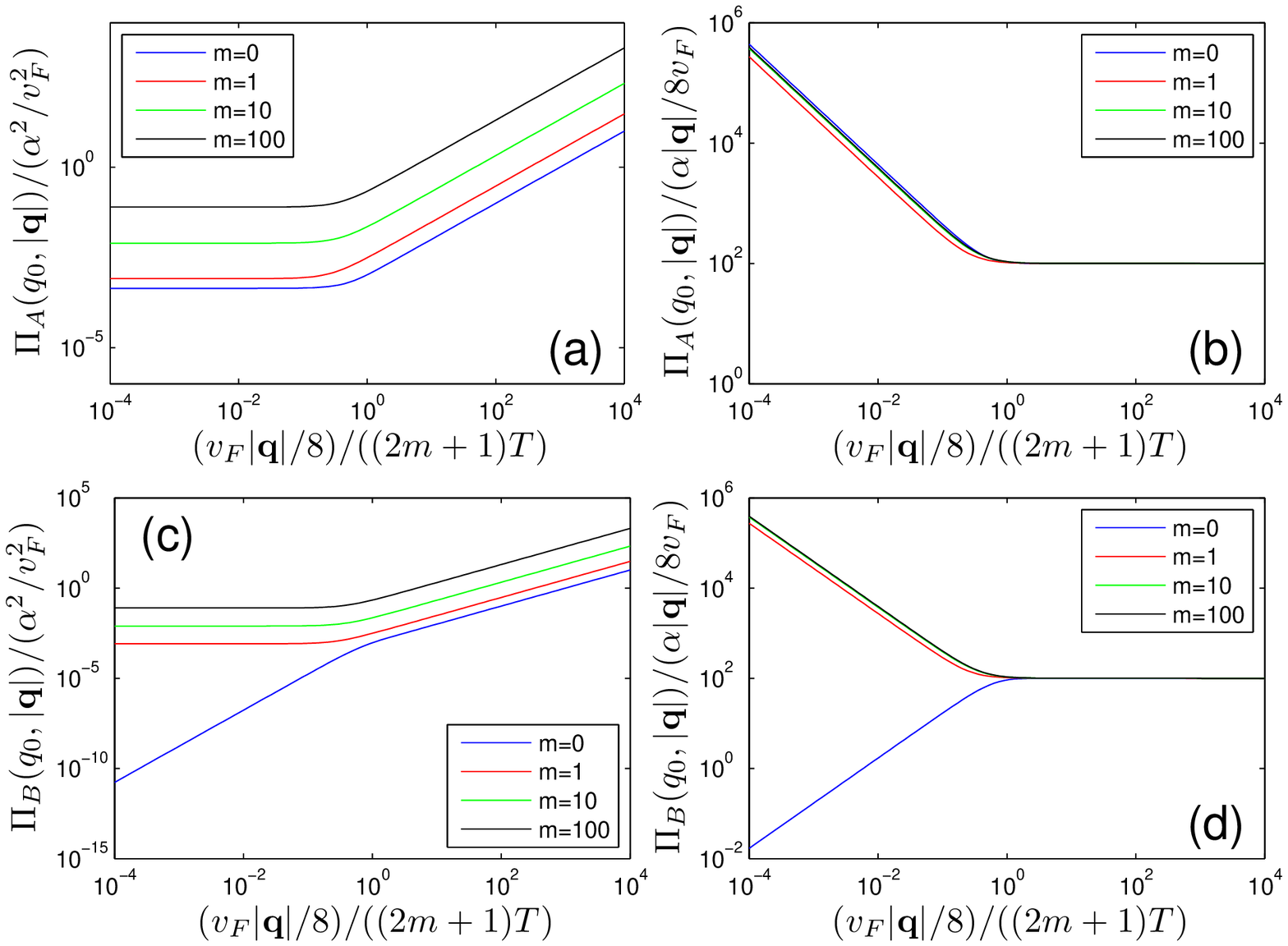}
\caption{Polarizations $\Pi(q_{0},|\mathbf{q}|)$ and
$\Pi_{B}(q_{0},|\mathbf{q}|)$ for different frequencies. Here, $q_0
= 2m\pi T$ and $T$ is chosen as $T/\alpha =
10^{-3}$.}\label{Fig:PolaFiniteFreB}
\end{figure}
with $\chi = \sqrt{x(1-x) \left(q_{0}^{2} +
v^{2}\mathbf{q}^2\right)}$. In the instantaneous approximation,
the energy dependence of the polarizations is dropped by
demanding $\Pi_{A,B}(q_0,\mathbf{q})\rightarrow
\Pi_{A,B}(q_0=0,\mathbf{q})$, which gives rise to
\begin{eqnarray}
\Pi_{A} = \frac{2\alpha T}{\pi v^{2}}
\int_{0}^{1}dx\ln\left[2\cosh\left(\frac{\sqrt{x(1-x)}
v|\mathbf{q}|}{2T}\right)\right], \nonumber
\end{eqnarray}
and
\begin{eqnarray}
\Pi_{B} = \frac{\alpha |\mathbf{q}|}{\pi
v}\int_{0}^{1}dx\sqrt{x(1-x)}
\tanh\left(\frac{\sqrt{x(1-x)}v|\mathbf{q}|}{2T}\right). \nonumber
\end{eqnarray}
To the leading order of $1/N$ expansion, the fermion self-energy is
given by
\begin{eqnarray}
\Sigma(p_{0},\mathbf{p}) &=& \frac{\alpha T}{N}\sum_{q_{0}}\int
\frac{d^2\mathbf{q}}{(2\pi)^2} \gamma_{\mu}
G(p_{0}-q_{0},\mathbf{p}-\mathbf{q})\gamma_{\nu} \nonumber \\
&& \times \Delta_{\mu\nu}(q_{0},\mathbf{q}).
\end{eqnarray}
The behavior of $\Sigma(p_{0},\mathbf{p})$ is mainly determined by
the low-energy properties of the gauge boson propagator $\Delta_{\mu
\nu}$, which in turn relies on the polarization functions $\Pi_A$
and $\Pi_B$. Before calculating $\Sigma(p_{0},\mathbf{p})$, it would
be helpful to first qualitatively analyze the properties of $\Pi_A$
and $\Pi_B$ at various values of $\mathbf{q}$.

As shown in Fig.~\ref{Fig:PolaFiniteFreA} and
Fig.~\ref{Fig:PolaFiniteFreB}, for the finite frequency components
of the gauge interaction, $q_{0}$ is considered as a small value in
the region $|\mathbf{q}| \gg q_{0}$, both
$\Pi_{A}(q_{0},|\mathbf{q}|)$ and $\Pi_{B}(q_{0},|\mathbf{q}|)$
approach the value $\frac{v|\mathbf{q}|}{8}$. In this region, the
self-energy corrections due to the longitudinal and transverse
components of gauge interaction should nearly cancel each other.
Therefore, the finite frequency components of the gauge interaction
will not induce singular fermion velocity renormalization in this
region. In the region $|\mathbf{q}| \ll q_{0}$, $q_{0}$ is a large
value and $q_{0}^{2}$ is an effective screening factor. Both
$\Pi_{A}$ and $\Pi_{B}$ approach to some finite values in this
region in the limit $|\mathbf{q}|\rightarrow0$, which implies that
the longitudinal and transverse components of gauge interaction are
both screened. In this case, the finite frequency components of
gauge interaction also cannot lead to singular velocity
renormalization.

For the zero frequency component of the gauge interaction, as shown
in Fig~\ref{Fig:PolaFiniteFreA} and Fig.~\ref{Fig:PolaFiniteFreB},
both $\Pi_{A}$ and $\Pi_{B}$ can be simplified to $|\mathbf{q}|/8$
if $|\mathbf{q}|> T$. Therefore, the fermion velocity is indeed not
renormalized at energy scales above $T$. At energy scales lower than
$T$, however, $T$ can be considered as a large variable and hence
the behavior of $\Pi_{A}$ becomes very different from that of
$\Pi_{B}$. In this region, we find that
\begin{eqnarray}
\Pi_{A}(0,\mathbf{q}) &\approx& \frac{2\alpha \ln2}{\pi}
\frac{T}{v^{2}}, \\
\Pi_{B}(0,\mathbf{q}) &\approx& \frac{\alpha}{12\pi}
\frac{\mathbf{q}^{2}}{T}.
\end{eqnarray}
Since $T$ is a relatively large quantity, now the longitudinal
component of gauge interaction is statically screened and does not
play an important role in the low energy region. Nevertheless, the
transverse component of gauge interaction remains long-ranged,
characterized by the fact that
\begin{eqnarray}
\lim_{\mathbf{q}\rightarrow 0}\Pi_{B}(0,\mathbf{q}) \rightarrow 0,
\end{eqnarray}
as required by the local gauge invariance. Therefore, the singular
contribution to the fermion self-energy can only be induced by the
zero frequency part of the transverse component of gauge
interaction. Taking advantage of this fact, we can simply ignore the
longitudinal component of gauge interaction and calculate the
fermion self-energy as follows:
\begin{widetext}
\begin{eqnarray}
\Sigma_{\mathrm{S}}(p_{0},\mathbf{p}) &=& \frac{\alpha
T}{N}\int\frac{d^2\mathbf{q}}{(2\pi)^2}
\gamma_{\mu}G(p_{0},\mathbf{p}-\mathbf{q})\gamma_{\nu}
\frac{B_{\mu\nu}}{\mathbf{q}^2+\Pi_{B}(\mathbf{q})}\nonumber
\\
&=&\frac{\alpha T}{N}\int\frac{d^2\mathbf{q}}{(2\pi)^2}
\gamma_{\mu}\frac{1}{p_{0} \gamma_{0} +v
\mathbf{\gamma}\cdot\left(\mathbf{p}-\mathbf{q}\right)}\gamma_{\nu}
\frac{\delta_{\mu i}\left(\delta_{ij} -
\frac{q_{i}q_{j}}{\mathbf{q}^{2}}\right)\delta_{j\nu}}{\mathbf{q}^2
+ \Pi_{B}(\mathbf{q})}\nonumber \\
&=& -\frac{\alpha T}{N}\int\frac{d^2\mathbf{q}}{(2\pi)^2}
\frac{p_{0}\gamma_{0} - v\mathbf{\gamma}\cdot(\mathbf{p}+\mathbf{q})
+ 2v \mathbf{\gamma}\cdot \mathbf{q}\frac{\mathbf{p} \cdot
\mathbf{q}}{\mathbf{q}^{2}}}{p_{0}^{2} \left(1 +
\frac{v^{2}\left(\mathbf{p} -
\mathbf{q}\right)^{2}}{p_{0}^{2}}\right)}
\frac{1}{\mathbf{q}^2+\Pi_{B}(\mathbf{q})}.
\end{eqnarray}
Since we are now considering the energy scales below $T$, as
explained above Eq.~(14), we can make the following approximations:
\begin{eqnarray}
\frac{1}{1+\frac{v^{2}\left(\mathbf{p}-\mathbf{q}\right)^{2}}{p_{0}^{2}}}
\approx 1-\frac{v^{2}\left(\mathbf{p} -
\mathbf{q}\right)^{2}}{p_{0}^{2}} \approx 1 +
\frac{2v^{2}\mathbf{p}\cdot\mathbf{q}-v^{2}\mathbf{q}^{2}}{p_{0}^{2}},
\end{eqnarray}
which is valid because $p_0 \propto T$. Now we can divide the
self-energy function into two parts:
\begin{eqnarray}
\Sigma_{\mathrm{S}}(p_{0},\mathbf{p}) &=&
-\frac{\alpha}{N}T\int\frac{d^2\mathbf{q}}{(2\pi)^2}
\frac{p_{0}\gamma_{0}-v\mathbf{\gamma} \cdot (\mathbf{p}+\mathbf{q})
+ 2v \mathbf{\gamma} \cdot \mathbf{q}
\frac{\mathbf{p}\cdot\mathbf{q}}{\mathbf{q}^{2}}}{p_{0}^{2}}
\left(1-\frac{v^{2}\mathbf{q}^{2}}{p_{0}^{2}}\right)
\frac{1}{\mathbf{q}^2+\Pi_{B}(\mathbf{q})}\nonumber
\\
&&-\frac{\alpha}{N}T\int\frac{d^2\mathbf{q}}{(2\pi)^2}
\frac{p_{0}\gamma_{0} - v\mathbf{\gamma}
\cdot(\mathbf{p}+\mathbf{q})+2v \mathbf{\gamma}\cdot \mathbf{q}
\frac{\mathbf{p}\cdot\mathbf{q}}{\mathbf{q}^{2}}}{p_{0}^{4}}
\frac{2v^{2}\mathbf{p}\cdot\mathbf{q}}{\mathbf{q}^2 +
\Pi_{B}(\mathbf{q})}.
\end{eqnarray}
\end{widetext}
Straightforward algebraic calculations show that
\begin{eqnarray}
\Sigma_{\mathrm{S}}(p_{0},\mathbf{p}) = \Sigma_{0}p_{0}\gamma_{0} +
\Sigma_{1}\mathbf{p}\cdot\mathbf{\gamma},
\end{eqnarray}
where
\begin{eqnarray}
\Sigma_{0}&=&-\frac{\alpha T}{N p_{0}^{2}}
\int\frac{d^2\mathbf{q}}{(2\pi)^2}
\frac{1}{\mathbf{q}^2+\Pi_{B}(\mathbf{q})}\nonumber
\\
&&+\frac{\alpha T}{N p_{0}^{4}}\int\frac{d^2\mathbf{q}}{(2\pi)^2}
\frac{v^{2}\mathbf{q}^{2}}{\mathbf{q}^2 + \Pi_{B}(\mathbf{q})},
\\
\Sigma_{1} &=& \frac{\alpha T}{N p_{0}^{4}} \int
\frac{d^2\mathbf{q}}{(2\pi)^2} \frac{v^{2}
\mathbf{q}^{2}}{\mathbf{q}^2 + \Pi_{B}(\mathbf{q})}.
\end{eqnarray}
The difference between $\Sigma_{0}$ and $\Sigma_{1}$ is given by
\begin{eqnarray}
\Sigma_{0}-\Sigma_{1} &\approx& -\frac{\alpha T}{Np_{0}^{2}}\int
\frac{d^2\mathbf{q}}{(2\pi)^2}\frac{1}{\mathbf{q}^2 +
\frac{\alpha}{12\pi} \frac{\mathbf{q}^2}{T}}.
\end{eqnarray}
To perform RG transformations, we need first to integrate over
momenta restricted in a thin shell of $[b\Lambda, \Lambda]$, where
$\Lambda$ is an ultraviolet cutoff and $b = e^{-l}$ with $l$ being a
varying length scale, which yields
\begin{eqnarray}
\Sigma_{0}-\Sigma_{1} &=& -\frac{\alpha}{2\pi^3 N(2n+1)^{2}\left(T +
\frac{\alpha}{12\pi}\right)}\int_{b\Lambda}^{\Lambda}
\frac{d|\mathbf{q}|}{|\mathbf{q}|}, \nonumber \\
&=& -\frac{\alpha}{2\pi^3 N(2n+1)^{2}\left(T +
\frac{\alpha}{12\pi}\right)}l.
\end{eqnarray}
The unusual velocity renormalization can be calculated from the
difference between $\Sigma_0$ and $\Sigma_1$ as follows
\begin{eqnarray}
\frac{d\ln v}{dl} = \frac{d\left(\Sigma_{0} - \Sigma_{1}\right)}{dl}
= -\frac{\alpha}{2\pi^3 N(2n+1)^{2}\left(T +
\frac{\alpha}{12\pi}\right)}.\nonumber \\
\end{eqnarray}
Solving this equation leads to the renormalized fermion velocity.
Based on the above calculations and analysis, we find that the
velocity depends on energy, momenta, and temperature approximately
as follows :
\begin{eqnarray}
v^{R}(p_0,\mathbf{p},T) = \left\{\begin{array}{ll}
\left(\frac{|\mathbf{p}|}{T}\right)^{\eta_{n}} &
|\mathbf{p}| < T, \\
1 & |\mathbf{p}| > T.
\end{array}\right.\label{Eq:VelocityRG}
\end{eqnarray}
The above expression shows that the originally constant velocity
acquires an anomalous dimension $\eta_n$:
\begin{eqnarray}
\eta_{n} = \frac{\alpha}{2\pi^3 N(2n+1)^{2}\left(T +
\frac{\alpha}{12\pi}\right)}.\label{Eq:EtaN}
\end{eqnarray}
in the low-energy region $|\mathbf{p}|< T$. Since $\eta_n
> 0$, the renormalized velocity $v^{R}(p_0,|\mathbf{p}|,T)$ vanishes
in the limit $|\mathbf{p}| \rightarrow 0$, which then leads to an
appropriate modification of the fermion dispersion. If we take the
zero temperature limit $T \rightarrow 0$, the velocity is simply
equal to unity, namely $v \equiv 1$, which is well expected since
QED$_3$ respects the Lorentz invariance at zero temperature.

We now examine the impact of the velocity renormalization. Since the
fermion dispersion is modified, it is reasonable to expect that many
physical quantities will be influenced, qualitatively or
quantitatively. From the recent research experience of graphene
\cite{Son07, Vafek07, Kotov, WangLiu12, WangLiu14} and high-$T_c$
superconductors \cite{Kim97, Huh08, Xu08, Liu12, She15}, we know
that unusual fermion velocity renormalization can lead to
significant changes of the spectral and thermodynamic properties of
massless Dirac fermions. It also strongly alters the critical
interaction strength for dynamical chiral symmetry breaking in
graphene \cite{WangLiu12}. Here, we consider one particular
quantity, namely the fermion specific heat, and leave the effects of
velocity renormalization on other physical properties to future
work.

For a (2+1)-dimensional non-interacting Dirac fermion system, the
specific heat is known to be proportional to $T^2$.  In the
following, we examine the influence of renormalized, $T$-dependent
fermion velocity on the specific heat. For simplicity, we first take
the zero-energy limit, and thus have $v^{R} \equiv
v^{R}(p_{0}=0,|\mathbf{p}|,T)$. In this limit, the corresponding
free energy is given by
\begin{eqnarray}
F(T) &=& -\frac{2NT}{\pi}\left[\int_{0}^{T}dk k
\ln\left(1+e^{-\left(\frac{k}{T}\right)^{\eta_{0}+1}}\right)\right.
\nonumber \\
&&\left. + \int_{T}^{+\infty}dk k \ln\left(1 + e^{-\frac{k}{T}}\right)\right]
\nonumber \\
&=& -\frac{2NT^3}{\pi}\left[\int_{0}^{1}dx x \ln
\left(1+e^{-x^{\eta_{0}+1}}\right)\right. \nonumber \\
&&\left. + \int_{1}^{+\infty}dx x\ln\left(1+e^{-x}\right)\right],
\end{eqnarray}
At $T \ll \alpha$, the anomalous dimension $\eta_0$ becomes
$T$-independent, i.e., $\eta_{0} \rightarrow \frac{6}{N\pi^2}$,
hence the corresponding specific heat is $C_{V}=-T\frac{\partial^{2}
F}{\partial T^{2}}\propto T^2$. To compute the free energy with
higher accuracy, we need to include the dependence of anomalous
dimension on both $p_0$ and $T$. At finite $T$, the energy $p_0$
takes a series of discrete values, which makes it difficult to do
analytic calculations. We therefore define the following mean value
of the renormalized fermion velocity $\bar{v}_{F}^{R}(\mathbf{p}) =
\left(\frac{|\mathbf{p}|}{T}\right)^{\bar{\eta}}$, where
$\bar{\eta}$ is obtained by performing an average over all the
frequencies:
\begin{eqnarray}
\bar{\eta} &=& \frac{\sum_{n=-\infty}^{+\infty}\eta_{n}
\frac{1}{(2n+1)^{2}}}{\sum_{n=-\infty}^{\infty}
\frac{1}{(2n+1)^2}}\nonumber \\
&=& \frac{1}{2N\pi}\frac{\alpha }{\pi^{2}
\left(T+\frac{2\alpha}{3\pi }\right)}
\frac{\sum_{n=-\infty}^{+\infty}
\frac{1}{(2n+1)^{4}}}{\sum_{n=-\infty}^{\infty}
\frac{1}{(2n+1)^2}}\nonumber \\
&=&\frac{1}{24N\pi}\frac{\alpha}{\left(T +
\frac{\alpha}{12\pi}\right)}.
\end{eqnarray}
Using the above expressions, we obtain the following averaged free
energy
\begin{eqnarray}
F_{\mathrm{avr}}(T) &=& -\frac{2NT^3}{\pi}\left[\int_{0}^{1}dx x \ln
\left(1 + e^{-x^{\bar{\eta}+1}}\right)\right. \nonumber \\
&&\left. + \int_{1}^{+\infty}dx x\ln\left(1+e^{-x}\right)\right].
\end{eqnarray}
Both analytical and numerical calculations show that the
corresponding specific heat $C_V(T)$ is still proportional to $T^2$
in the low temperature regime, but its coefficient is strongly
altered by the anomalous dimension.

We now remark on the issue of gauge invariance. In a quantum gauge
field theory, it is of paramount importance to obtain a
gauge-independent quantity, which, however, is a highly nontrivial
task. The studies of QED$_{3}$ have also been suffering from this
problem for three decades. In Ref.~\cite{Appelquist88}, Appelquist
\emph{et al.} utilized the Landau gauge to construct DSE for
dynamical fermion mass and found a finite critical fermion flavor
$N_{c} = \frac{32}{\pi^2}$ to the lowest order of $1/N$ expansion.
Subsequent work of Nash \cite{Nash89} included the impact of the
next-to-leading order correction and claimed to obtain a
gauge-independent critical flavor $N_{c} =
\frac{4}{3}\frac{32}{\pi^2}$. More recently, Fischer \emph{et al.}
\cite{Fisher04} studied DCSB by analyzing the self-consistently
coupled DSEs of fermion and gauge boson propagators. An
\emph{ansatz} for the vertex correction was introduced in
Ref.~\cite{Fisher04} to fulfill the Ward-Green-Takahashi identity.
In the Landau gauge, they found the critical flavor $N_{c}\approx
4$, which is close to the value of Appelquist \emph{et al.}
\cite{Appelquist88}. However, after comparing the results obtained
in various gauges, they showed that the conclusion is apparently not
gauge invariant. Certainly, one would obtain a gauge independent
conclusion if the full DSEs were solved without making any
approximations. This is practically not possible and it is always
necessary to truncate the complicated DSEs in some proper way. How
to truncate the DSEs in a correct way so as to get gauge invariant
results is still an open question \cite{Bashir08, Bashir09,
Goecke09, Lo11}. The same problem is encountered in the application
of QED$_{3}$ \cite{Rantner, Franz, Khveshchenko02A, Khveshchenko02B,
Ye03, Gusynin03, Khveshchenko03, Franz03} to interpret some
interesting experimental facts of cuprate superconductors
\cite{Ding96, Feng99, Valla00}. In this case, it remains unclear how
to obtain a gauge independent propagator for the massless Dirac
fermions \cite{Gusynin03, Khveshchenko03, Franz03}.

In the above discussions, we have used the Landau gauge, which is
widely used in the studies of QED$_{3}$ and expected to be the most
reliable gauge \cite{Bashir08, Goecke09, Lo11}. If we include an
arbitrary gauge parameter $\xi$, the effective gauge boson
propagator becomes
\begin{eqnarray}
\Delta_{\mu\nu}(q_{0},\mathbf{q}) &=& \frac{A_{\mu\nu}}{q_{0}^2
+\mathbf{q}^{2}+\Pi_{A}(q_{0},\mathbf{q})} \nonumber \\
&&+\frac{B_{\mu\nu}}{q_{0}^{2} + \mathbf{q}^2 +
\Pi_{B}(q_{0},\mathbf{q})} + \xi\frac{q_{\mu}q_{\nu}}{q^4}.
\end{eqnarray}
After analogous RG calculations, we find that the anomalous
dimension receives an additional term:
\begin{eqnarray}
\eta_{n}' = \eta_{n}+\eta_{\xi},
\end{eqnarray}
where
\begin{eqnarray}
\eta_{\xi} = \frac{e^2 \xi}{2\pi(2n+1)^2 T}.
\end{eqnarray}
It appears that the anomalous dimension and thus the renormalized
velocity depends on the gauge parameter $\xi$. We expect this gauge
dependence can be removed if higher order corrections could be
properly incorporated. Technically, computing higher order
corrections to fermion self-energy in finite-$T$ QED$_{3}$ is much
harder than zero-$T$ QED$_{3}$ since the summation over discrete
frequency and integration of momenta have to be performed
separately.

Though being gauge dependent, we still believe that our RG results
are qualitatively correct. To gain a better understanding of the
essence of singular velocity renormalization and the appearance of
anomalous dimension, we now make a comparison between a number of
physically similar systems. The first example is zero-$T$ QED$_{3}$
at a finite chemical potential $\mu$, which induces a finite Fermi
surface of Dirac fermions. The Fermi surface explicitly breaks the
Lorentz invariance and also leads to static screening of the
longitudinal component of gauge interaction. The transverse
component of gauge interaction is still long ranged and thus is able
to generate singular velocity renormalization. It was previously
shown in Ref.~\cite{Wang12} that the velocity behaves like
$v^{R}\propto(\frac{k}{\mu})^{\eta}$, where $\eta$ is a finite
number. The second example is graphene in which massless Dirac
fermions emerge as low-energy excitations. The long Coulomb
interaction also breaks Lorentz invariance explicitly, and is
unscreened due to the vanishing of of zero-energy density of states.
In this case, the fermion velocity is singularly renormalized and
increases indefinitely as the energy is lowering \cite{Gonzalez94,
Gonzalez99, Son07}. As aforementioned, analogous velocity
renormalization takes place in the effective QED$_{3}$ theory of
high-$T_c$ cuprate superconductors \cite{Lee06, Kim97, Kim99} and
also at a nematic quantum critical points \cite{Huh08, Wang11,
Liu12, She15} which are also resulting from the breaking of Lorentz
invariance. We can extract a generic principle from all these
examples that the long-range interaction always leads to singular
fermion velocity renormalization once the Lorentz invariance is
broken. It is known that the Lorentz invariance is broken at finite
$T$ in QED$_{3}$ \cite{Dorey92, Lee98, Triantaphyllou}. According to
this principle, the fermion velocity has to be singularly
renormalized. Therefore, our RG results for the renormalized
velocity and the anomalous dimension should be qualitatively
reliable, though quantitatively not precise due to the gauge
dependence.

Recently, three-dimensional (3D) Dirac semimetal state was observed
at the quantum critical point between a bulk topological insulator
and a trivial band insulator \cite{Xu11}. Experiments also confirmed
that Na$_{3}$Bi \cite{Liu14} and Cd$_{3}$As$_{2}$ \cite{Neupane14}
are 3D Dirac semimetals in which the massless Dirac fermions are
stable due to the protection of crystal symmetry. Isobe and Nagaosa
\cite{Isobe12, Isobe13} showed that in the presence of an
electromagnetic field, the velocity of Dirac fermions does not
receive singular renormalization but flows to some finite value in
the lowest energy limit, which is a consequence of the emergence of
Lorentz invariance. However, if a finite chemical potential is
induced in 3D Dirac semimetals by doping, the longitudinal component
of electromagnetic field will be screened. However, the transverse
component of electromagnetic field is not screened and is able to
result in singular renormalization of fermion velocity. Therefore,
the doped 3D Dirac semimetals placed in an electromagnetic field
provides an ideal platform for measuring singular fermion velocity
renormalization.

We next would like to connect our analysis to the issue of infrared
divergence. In the ordinary calculations based on perturbation
expansion or non-perturbative DSEs of fermion self-energy, the lower
limit of momenta is zero. At finite $T$, there is an infrared
divergence in the fermion self-energy induced by the zero frequency
part of the transverse component of gauge interaction \cite{Lee98,
Lo11B, WangLiuZhang15}. As pointed out by Lo and Swanson \cite{Lo11B},
this divergence has not been seriously considered in the previous
studies, where this problem is usually bypassed by completely
ignoring the transverse component of gauge interaction. They showed
\cite{Lo11} that this infrared divergence is endemic in finite-$T$
QED$_{3}$ and proposed to remove it by choosing a proper
$T$-dependent gauge parameter. This strategy is essentially
equivalent to dropping the zero frequency part of the transverse
component of gauge interaction but retaining the non-zero
frequencies. In the modern RG theory \cite{Shankar94}, one needs to
integrate over field operators defined in a thin momentum shell
$(b\Lambda,\Lambda)$. After performing RG manipulations, there will
be a singular renormalization for some quantities, such as fermion
velocity, caused by the long-range interaction. This singular
renormalization should have important influence on the infrared
behaviors of QED$_{3}$. It would be interesting and also challenging
to study whether the infrared divergence appearing in the DSE of
dynamical fermion mass \cite{Lee98, Lo11B, WangLiuZhang15} can be
eliminated by taking into account the influence of singular velocity
renormalization.

In summary, we have studied the renormalization of Dirac fermion
velocity in QED$_3$ at finite temperatures by means of RG method. We
first demonstrate that the velocity renormalization is a consequence
of the explicit breaking of Lorentz invariance due to thermal
fluctuations. We then obtain the renormalized fermion velocity as a
function of energy, momentum, and temperature, as shown in
(\ref{Eq:VelocityRG}) and (\ref{Eq:EtaN}). We have also computed the
specific heat after taking into account the velocity
renormalization. It would be interesting to further study its
impacts on DCSB \cite{Dorey92, WangLiuZhang15} and non-Fermi liquid
behaviors \cite{WangLiu10A, WangLiu10B} in the future. Moreover, we
emphasize that the velocity renormalization can be testified by
realistic experiments. Actually, recent experiments have already
extracted the detailed momentum dependence of renormalized fermion
velocity (caused by long-range Coulomb interaction between Dirac
fermions) in graphene \cite{Elias11, Yu13, Siegel12}. Since
QED$_{3}$ is widely believed to be the effective field theory of a
number of condensed matter systems \cite{Lee06, Affleck, Kim97,
Kim99, Rantner, Franz, Herbut, Liu02, Liu03, Ran07, Hermele08,
Gusynin04, Gusynin07, Raya08, Klebanov, Lu14, Metlitski15, Wang15,
Mross15}, it would be possible to probe the predicted unusual
velocity renormalization in certain angle resolved photoemission
spectroscopy experiments \cite{Siegel12}.

We acknowledge financial support by the National Natural Science
Foundation of China under Grants No.11504379, No.11574285, and
No.U1532267.

\end{document}